\documentclass[twocolumn,superscriptaddress,showpacs,prl,amsmath,amssymb]{revtex4}
\usepackage{graphicx}
\usepackage{bm}
\begin{document}

\title{Survival in equilibrium step fluctuations}

\author{ C. Dasgupta}
\altaffiliation{Permanent address: Department of Physics, Indian
Institute of Science, Bangalore 560012, India
}
\affiliation{
Condensed Matter Theory Center,
Department of Physics, University of Maryland, College Park, Maryland 20742-4111
}
\author{ M. Constantin}
\affiliation{
Condensed Matter Theory Center,
Department of Physics, University of Maryland, College Park, Maryland 20742-4111
}
\affiliation{Materials Research Science and Engineering Center,
Department of Physics, University of Maryland, College Park, MD 20742-4111
}
\author{ S. \surname{Das Sarma}}
\affiliation{
Condensed Matter Theory Center,
Department of Physics, University of Maryland, College Park, Maryland 20742-4111
}
\author{Satya N. Majumdar}
\affiliation{Laboratoire de Physique Quantique, UMR C5626, Universite Paul
Sabatier, 31062 Toulouse Cedex, France}

\begin{abstract}
We report the results of analytic and numerical investigations of the
time scale of survival or non-zero-crossing probability $S(t)$ in equilibrium
step fluctuations described by Langevin equations appropriate for
attachment/detachment and edge-diffusion limited kinetics. An exact
relation between long-time behaviors of the survival probability and
the autocorrelation function is established and numerically verified.
$S(t)$ is shown to exhibit simple scaling behavior as a function of
system size and sampling time. Our theoretical results are in agreement
with those obtained from an analysis of experimental dynamical STM data
on step fluctuations on Al/Si(111) and Ag(111) surfaces.
\end{abstract}


\maketitle

There has been much recent theoretical \cite{krug,us1} and
experimental \cite{us1,expt2} interest in the persistence behavior
of fluctuating steps on a vicinal surface. A persistence probability
$P(t)$, defined as the probability that the position (``height'') of
the step edge at a point along a fluctuating step {\it does not return to
its initial value (at time $t=0$)} over time $t$ is
found \cite{krug,us1,expt2} in these studies to decay in time as a
power-law, $P(t) \propto t^{-\theta}$, for large $t$, where $\theta$
is the so-called {\it persistence exponent}. Similar power-law
behavior of the persistence probability has also been found in
experiments \cite{exp} for other physical processes.
It turns out that the precise definition of $P(t)$ is absolutely
crucial for the power-law behavior discussed in the recent surface
fluctuations literature. If, instead of considering the 
probability of not returning to the initial position, one defines a
{\it survival probability}, $S(t)$, as the probability of the
dynamical step height (at a fixed but arbitrary spatial location)
not returning in time $t$ to its average (``equilibrium'') level,
then, quite surprisingly, it was found in a recent experimental
study \cite{us1} of thermal fluctuations of surface steps that
$S(t)$ actually manifests, in sharp contrast to the power-law
behavior of $P(t)$, an exponential decay,
$S(t)\propto \exp(-t/\tau_s)$, at long times, where $\tau_s$
is the survival time scale. This exponential behavior of $S(t)$
has remained theoretically unexplained.

In this report we provide a definitive theoretical explanation 
for this exponential temporal behavior of the surface fluctuation
survival probability using rigorous (analytical) arguments and direct
(numerical) simulations. Survival and persistence turn out 
to be {\it identical} in problems related to Ising spin 
dynamics, where one is interested in the probability 
that a spin has not changed its sign (has not ``flipped'')
up to time $t$ \cite{ft2}. This is due to the {\it discrete}
nature of Ising spin dynamics, where a spin flip ensures a
change of sign with respect to both the initial and the average
(or any other reference) value of the stochastic variable.
In contrast, the {\it continuous} nature of surface fluctuation
dynamics, where the step height is a continuous variable, leads
to a fundamental qualitative difference between $P(t)$ and $S(t)$. This
qualitative difference was noted as an experimental fact and an unsolved
puzzle in Ref.~\cite{us1} for equilibrium step fluctuations on
Al/Si(111) \cite{ft3}. We emphasize that the probabilities
$P(t)$ and $S(t)$ provide completely different physical information
about surface step fluctuations: while persistence characterizes the
universality class of the dynamical process through the persistence
exponent $\theta$, survival, as discussed in this paper, 
provides useful information about the physical mechanisms 
(and their characteristic time scales) underlying step 
fluctuations in the long-time limit.

Another time scale that invariably enters experimental and numerical
measurements of any statistical quantity is {\it the sampling time}
$\delta t$ (the interval between successive measurements of the
step position). An understanding of the effects \cite{samp} of
a finite $\delta t$ on the measured survival probabilities
is necessary for comparing experimental and numerical results
with theoretical predictions.

In this paper, we present the results of a detailed study of the
behavior of $S(t)$ for two linear Langevin equations that
describe \cite{bartelt,rev2} step fluctuations under
attachment/detachment (``high-temperature'') and edge-diffusion
(``low temperature'') limited kinetics. We first show analytically
that if the equilibrium autocorrelation function $C(t)$ of height
fluctuations decays exponentially at long times, then $S(t)$
must also decay exponentially with a time scale that is proportional
to the correlation time (the time scale of the decay of $C(t)$).
This prediction is verified from numerical simulations
of the Langevin equations. The simulation results also provide
information about the dependence of the measured survival probability 
on the sampling time $\delta t$. We show that the survival
probability $S(t,L,\delta t)$ exhibits simple scaling
behavior in both models. Finally, we use available
experimental data \cite{us1,expt2,expt3} to calculate 
$C(t)$ and $S(t)$ for two physical systems that are believed 
to be described by these two Langevin equations and show that 
the experimental results are consistent with our predictions.

High-temperature step fluctuations dominated by atomistic attachment
and detachment at the step edge are known~\cite{bartelt,rev2} to be well
described by the second-order non-conserved linear Langevin equation
\begin{equation}
\frac {\partial h(x, t)} {\partial t} =
\frac{\Gamma_a \tilde{\beta}}{k_BT} \frac{\partial^{2} h(x,t)}{\partial x^2} +
\eta(x, t).
\label{eweqn}
\end{equation}
Here, $h(x,t)$ is the dynamical height fluctuation (position of the step
edge measured from its equilibrium value) at lateral point $x$ along the
step and time $t$, $\Gamma_a$ is the ``step mobility'', $\tilde{\beta}$ is the
step-edge stiffness, and $\eta(x,t)$ is a nonconserved Gaussian noise
satisfying $\langle \eta(x, t) \eta(x^{\prime},t^{\prime}) \rangle
= 2 \Gamma_a \delta(x-x^{\prime}) \delta(t-t^{\prime})$.
Low-temperature step fluctuations dominated by the step edge diffusion
mechanism are, on the other hand, described by the fourth order conserved
Langevin equation
\begin{equation}
\frac {\partial h(x, t)} {\partial t} =
-\frac{\Gamma_h \tilde{\beta}}{k_BT} \frac{\partial^{4} h(x,t)}{\partial x^4} +
\eta_c(x, t),
\label{conseqn}
\end{equation}
with $\langle \eta_{c}(x,t) \eta_{c}(x^{\prime},t^{\prime}) \rangle =
-2 \Gamma_h \nabla_x^2 \delta(x-x^{\prime})
\delta(t-t^{\prime})$.

Space- and time-dependent correlation functions of height fluctuations in
these two {\it linear} equations may be calculated~\cite{bartelt,rev2}
easily by Fourier transforms. We assume that
height fluctuations are measured from the spatial
average of $h(x,t)$, so that the $k=0$ Fourier component of $h(x,t)$ is zero at
all times. The autocorrelation function of $\tilde{h}(k,t)$, the Fourier
transform of $h(x,t)$, has the following form in the long-time equilibrium
state:
\begin{equation}
\langle \tilde{h}(k,t_1)\tilde{h}(-k,t_2)\rangle =
\frac{k_BT}{\tilde{\beta}k^2} \exp(-\Gamma \tilde{\beta} k^z |t_1-t_2|/k_BT),
\label{kcorr}
\end{equation}
where $z=2$, $\Gamma=\Gamma_a$ for Eq.(\ref{eweqn}), and $z=4$,
$\Gamma=\Gamma_h$ for
Eq.(\ref{conseqn}). The autocorrelation function of height fluctuations at
equilibrium is then given by
\begin{eqnarray}
C(t) &=& \langle h(x,t_{1}) h(x,t_{2}) \rangle \nonumber \\
&=& \frac{2k_BT}{\tilde{\beta}}
\int_{k_{min}}^\infty \frac{dk}{2\pi}\frac{\exp(-\Gamma \tilde{\beta} k^z
|t|/k_BT)}{k^2}
\label{xcorr}
\end{eqnarray}
where $t=t_1-t_2$ and $k_{min} = 2\pi/L$ for a finite system of linear
dimension $L$. This implies that $C(t)$ exhibits
an exponential decay at long times, $C(t) \propto \exp(-t/\tau_c)$, where the
correlation time $\tau_c$ is equal to
$k_B T L^z/(2\pi)^z \Gamma \tilde{\beta}$.

There are other physical mechanisms that may lead to an exponential decay of
$C(t)$ at long times. The fluctuations of a particular step are affected by
its interaction with neighboring steps. These interaction effects,
which are negligible at relatively short time scales if the spacing
between neighboring steps is large, may become important at long
times. If one assumes that the step fluctuates in a harmonic 
confining potential \cite{bartelt}, $(\lambda/2)\int h^2(x) dx$, 
then one obtains an additional term, $- \lambda \Gamma_a h(x,t)/k_BT$, 
in the right-hand of Eq.(\ref{eweqn}), and 
$\lambda \Gamma_h \nabla_x^2 h(x,t)/k_B T$ 
in Eq.(\ref{conseqn}). The function $C(t)$ is then given by
\begin{equation}
C(t) = 2 k_B T
\int_{k_{min}}^\infty \frac{dk}{2\pi}\frac{\exp[-\Gamma (\tilde{\beta} k^z
+\lambda k^{z-2})|t|/k_BT)]}{\tilde{\beta}k^2+\lambda}.
\label{corrn}
\end{equation}
This, again, leads to an exponential decay of $C(t)$ at long times, with the
correlation time $\tau_c$ a function of $\lambda$ and $L$.

There exists a rigorous theorem \cite{pers} that states that if the
autocorrelation function $C(t)$ of a stationary Gaussian process decays
exponentially in time, then its survival probability $S(t)$ must 
also decay exponentially for large $t$, $S(t) \propto \exp(-t/\tau_s)$,
with the survival time scale $\tau_s$ proportional to the
correlation time $\tau_c$. The constant of proportionality,
$c \equiv \tau_s/\tau_c$, which must be less than unity and
independent of the system size $L$, is usually nontrivial,
being determined by the full functional form of $C(t)$. Since the
height fluctuations in our models represent a stationary Gaussian process at
equilibrium, this rigorous result applies for the survival probability
of these fluctuations. Thus, we arrive at a very general, exact
result that the survival probability of equilibrium step
fluctuations should decay exponentially at long times if the autocorrelation
function does so. This is the first important result of this article.
Further, measurements of the ratio $c$ of the two time scales may provide
valuable information about the nature of the processes involved in the step
fluctuations.

We have investigated these aspects in a detailed numerical study in
which a simple Euler scheme \cite{krug} is used to numerically integrate
spatially discretized versions of Eqs.(\ref{eweqn}) and
(\ref{conseqn}). All the results reported here were obtained in 
the equilibrium regime. $S(t)$ is measured as the probability 
that the height fluctuation $h_i$ at a particular site $i$ does 
not cross the average step height (which is conveniently chosen as 
the ``zero'' of the height stochastic variable) over time $t$, 
averaged over all sites and many ($\sim 10^4-10^6$) independent runs. 
$C(t)$ is calculated exactly using discretized versions of 
Eqs.(\ref{xcorr}) and (\ref{corrn}).

\begin{figure}
\includegraphics[height=7cm,width=8.5cm]{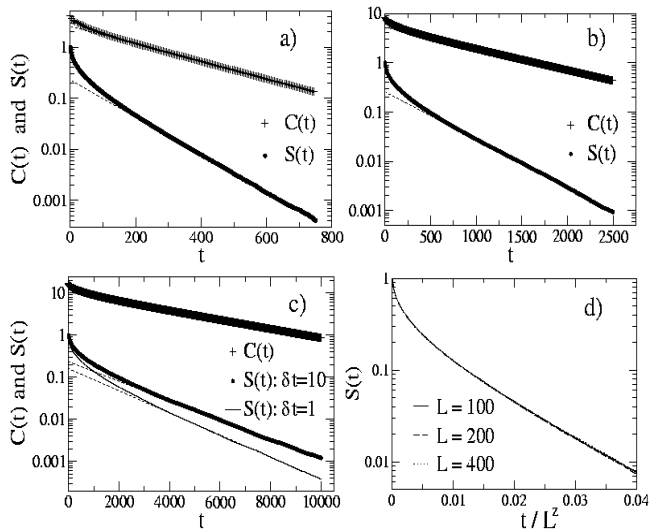} 
\caption{\label{fig1} $S(t)$ and $C(t)$ for the Langevin equation of
Eq.(1). The dashed lines are fits of the long-time data to an
exponential form. In panels (a-c), the uppermost plots show
the data for $C(t)$. Panel (a): $L=100$, $\delta t=0.625$. Panel (b):
$L=200$, $\delta t=2.5$. Panel (c): $L=400$, $\delta t=10.0$ (upper plot) and
$\delta t=1.0$ (lower plot).
Panel (d): Finite-size scaling of $S(t,L,\delta t)$.
Results for $S$ for 3 different sample sizes with 
the same value of $\delta t/L^z$ ($z=2$) are plotted versus $t/L^z$.}
\end{figure}

\begin{figure}
\includegraphics[height=7cm,width=8.5cm]{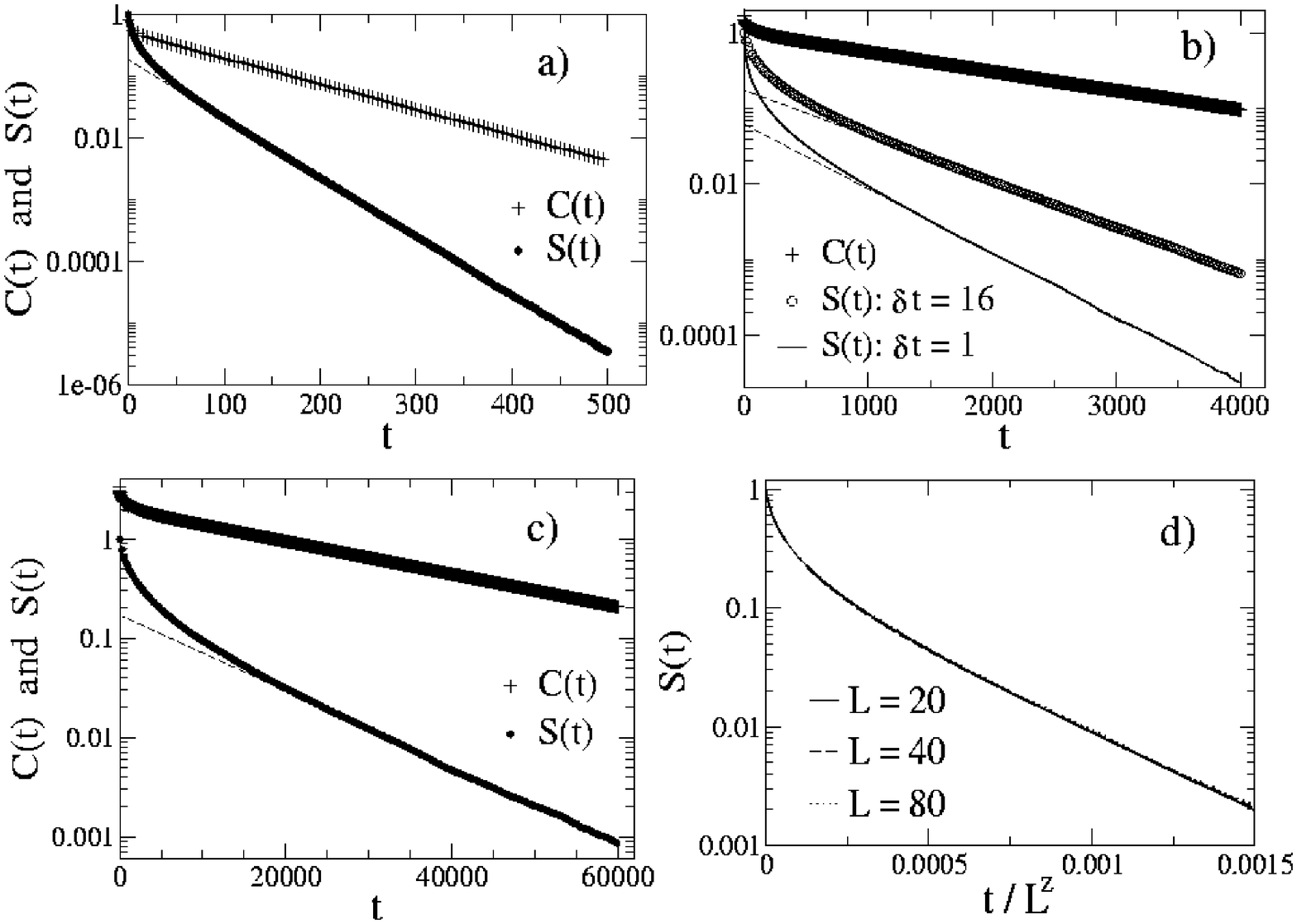} 
\caption{\label{fig2} $S(t)$ and $C(t)$ for the Langevin equation 
of Eq.(2). In panels (a-c), the uppermost plots show the results 
for $C(t)$. The dashed lines are fits of the long-time data to an 
exponential form. Panel (a): $L=20$, $\delta t=1$. Panel (b):
$L=40$, $\delta t=16$ (upper plot) and $L=40$, $\delta t=1$ (lower plot).
Panel (c): $L=80$, $\delta t=256$. Panel (d): Finite-size scaling of
$S(t,L,\delta t)$. Results for $S$ for 3 different sample sizes
with the same value of $\delta t/L^z$ ($z=4$) are
plotted versus $t/L^z$.}
\end{figure}

Typical results for $C(t)$ and $S(t)$ are shown in Fig.\ref{fig1} (a-c) for
Eq.(\ref{eweqn}) and in Fig.\ref{fig2} (a-c) for
Eq.(\ref{conseqn}) [$\lambda = 0$ in both cases].
As indicated in the figures, we used several different
values of the sampling time $\delta t$ in the measurement of $S(t)$
[$C(t)$ is, of course, independent of $\delta t$]. It is clear from the
plots that both $C(t)$ and $S(t)$ decay exponentially at long times.
The time scales $\tau_c$ and $\tau_s$ are extracted from exponential fits
shown in the semi-log plots as dashed straight lines. The dependence
of $\tau_c$ on $L$ is given exactly by $\tau_c(L)= (L/2\pi)^z$. The
results for different values of $L$ shown in Figs. \ref{fig1} and
\ref{fig2} indicate that $\tau_s$ also increases rapidly as $L$ is
increased. However, we find that the calculated values of $\tau_s$
extracted from the $S(t)$ data obtained for different $L$ using
{\it the same sampling time $\delta t$} exhibit small but
clear deviations from the expected proportionality to $L^z$.
These small deviations result from a weak dependence of $S(t)$ on
the sampling time $\delta t$. As shown
in Fig.\ref{fig1}c and Fig.\ref{fig2}b, the rate of the exponential decay of
$S(t)$ at large $t$ depends weakly on the value of $\delta t$ used in the
measurement of $S$. This is in accordance with the analytic predictions of
Ref.\cite{samp}.  Since the only time scale in the problem is $\tau_c$ (as
mentioned above, $\tau_s$ should be proportional to $\tau_c$), the dependence
of $S(t)$ on the sampling time $\delta t$ should involve the scaling
combination $\delta t/\tau_c$.
Since $\tau_c(L) \propto L^z$ in our models, this argument suggests that
the sampling time should be chosen to be proportional to $L^z$
if the survival probabilities for different values of $L$ are
to be tested for scaling. Indeed, as shown in Fig.\ref{fig1}d,
the values of $S(t)$ obtained for $L$ = 100, 200 and 400
using $\delta t$ = 0.625, 2.5 and 10.0, respectively,
(so that $\delta t \propto L^z$ with $z=2$) all fall on the same
scaling curve when plotted as functions of $t/L^z$ with $z=2$.
As shown in Fig.\ref{fig2}d, a similar scaling collapse is obtained
for Eq.(\ref{conseqn}). Here, the sampling
times for different $L$ are chosen to be proportional to $L^4$, and the
best scaling collapse is obtained when the data for different $L$ are plotted
against $t/L^z$ with $z \simeq 3.95$. These results
establish that the {\it full function $S(t,L,\delta t)$} (not just the
asymptotic long-time part) has the scaling form
\begin{equation}
S(t,L,\delta t) = f(t/L^z,\delta t/L^z),
\label{scaling}
\end{equation}
where the function $f(x,y)$ decays exponentially for large values of $x$ and
the rate of this decay increases slowly as $y$ is decreased. This finite-size
scaling behavior of $S$, which represents the second important result
of our study, is similar to that found \cite{manoj} for the
persistence probability in a coarsening system. However, the 
dependence on the sampling time, $essential$ in our scaling
considerations, was not analyzed in Ref.\cite{manoj}. 

We have also studied the behavior of $C(t)$ and $S(t)$ for Eq.(\ref{eweqn})
when the value of $\tau_c$ is primarily determined by the presence of a
nonzero $\lambda$ associated with step-step interaction
(cf. Eq.~(\ref{corrn}) above).
By varying $\lambda$ and $\delta t$ for a system with $L=400$, we find
that $S(t,\lambda,\delta t)$ exhibits excellent scaling behavior as a
function of $t/\tau_c$ if the quantity $\delta t/\tau_c$ is
held constant. Therefore, we conclude that $S$ is a function of the scaling
variables $t/\tau_c$ and $\delta t/\tau_c$, irrespective of the origin of the
finite value of the correlation time $\tau_c$.

For $\lambda=0$, the ratio $c=\tau_s/\tau_c$ for Eq.(\ref{eweqn})
decreases from about 0.57
to about 0.41 as the ratio $\delta t/\tau_c$ is decreased from 0.025 to
$2.5\times 10^{-4}$, indicating that $c \simeq 0.4$ in the
$\delta t \to 0$ limit. For
relatively large $L$ and nonzero $\lambda$ where $\tau_c$ is determined
primarily by the value of $\lambda$, we find that $c \simeq 0.47$ for
$\delta t/\tau_c = 0.025$. The difference between the values of $c$
for the same value of $\delta t/\tau_c$ in the two
cases reflects the expected dependence
of $c$ on the details of $C(t)$. For Eq.(\ref{conseqn}) with $\lambda=0$, the
value of $c$ decreases from about 0.44 to about 0.30 as $\delta t/\tau_c$ is
decreased from 0.01 to $6\times 10^{-4}$. The qualitative behavior of $c$
as a function of $\delta t/\tau_c$ is similar in all the cases we have
considered, and is consistent with the general predictions of Ref.\cite{samp}.
\begin{figure}
\includegraphics[height=3.8cm,width=8.5cm]{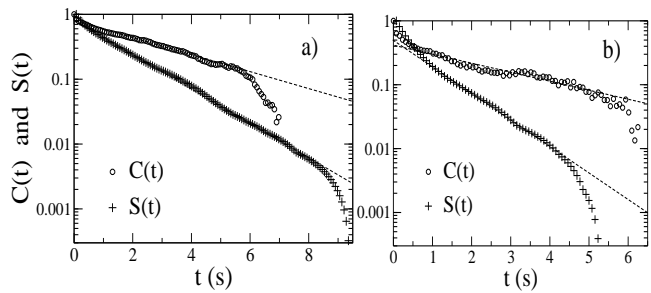} 
\caption{\label{fig3} $S(t)$ and $C(t)$ for two experimental 
systems. The dashed lines are fits of the long-time data to 
an exponential form. Panel (a): Al/Si(111) at $T$ = 970K.
Panel (b): Ag(111) at $T$ = 450K.}
\end{figure}

We have also used dynamical STM data to calculate  $C(t)$ and $S(t)$ for
two experimental systems: Al/Si(111) at relatively high temperatures, which
is believed~\cite{us1,lyubin,lyubin2} to provide a physical realization
of Eq.(\ref{eweqn}), and Ag(111) at relatively low temperatures where the
step fluctuations are expected~\cite{expt2,expt3} to be governed by the
conserved Eq.(\ref{conseqn}). Some of the results of this analysis are shown
in Fig.\ref{fig3}. For Al/Si(111) at 970K we find exponential decay of both
$C(t)$ and $S(t)$. The value of the ratio $c$ obtained from the estimates
of $\tau_c$ and $\tau_s$ is close to 0.5. This value is right in the middle
of the range of values of $c$ obtained from our numerical study of
Eq.(\ref{eweqn}). The Ag(111) at 450K data is characterized by 
$c \simeq 0.34$, which is again in the range of values obtained 
in the numerical study of Eq.(\ref{conseqn}). We, therefore, 
conclude that the available experimental data on $S(t)$ 
and $C(t)$ are consistent with our theoretical results.

Experimental data on the Al/Si(111) system are available at several
temperatures between 770K and 1020K. As reported in Ref.\cite{us1}, $S(t)$
decays exponentially at long times at all these temperatures, with $\tau_s$
decreasing from 3.6s to 0.9s as $T$ is increased from 770K to 1020K. Using
these values of $\tau_s$ (actually, the corresponding values of $\tau_c$
obtained from the relation $\tau_s/\tau_c \simeq 0.5$) together with the
values of the parameters $\Gamma$ and $\tilde{\beta}$ obtained from other
measurements \cite{lyubin,lyubin2}, we have calculated an ``effective length''
$L_{eff}$ that would lead to the observed finite value of $\tau_c$ if it
resulted from a finite length of the sample. The value of $L_{eff}$
is found to decrease from $4020\AA$ to $389\AA$ as $T$ is decreased from
1020K to 770K. These values are much smaller than the nominal step
lengths in the experimental sample. The observed $T$-dependence of $L_{eff}$
is inconsistent with the possibility that the finite values of $\tau_c$ are
due to a nonzero value of the parameter $\lambda$: the length scale associated
with $\lambda$ should increase~\cite{bartelt} as $T$ is decreased. It is
possible that $L_{eff}$ is a measure of the typical length of a step edge
between adjacent points that are held fixed by some kind of pinning centers.
Since pinning becomes more effective at low $T$, this mechanism would
provide a qualitative explanation of why $L_{eff}$ decreases as $T$ is
reduced. Yet another possibility is that $L_{eff}$ is a measure
of the length scale over which step edge fluctuations are effectively
equilibrated. 

To conclude, we have shown analytically and numerically
that the survival probability of equilibrium step fluctuations
on vicinal surfaces decays exponentially at long times, and
have established a relation between the time scales characterizing
the exponential decay of the survival probability and the
autocorrelation function. Our theory explains the puzzling
experimental finding of an exponential decay of $S(t)$
reported in Ref.~\cite{us1}. We have also shown that
the survival probability exhibits simple scaling as a
function of the system size and the sampling time, which plays
a very important role in the measurement of $S(t)$.

The authors gratefully acknowledge discussions with E.D. Williams,
O. Bondarchuk and D.B. Dougherty. S.N.M. thanks A. Bray for
useful discussions. This work is partially supported by US-ONR and
NSF-DMR-MRSEC at the University of Maryland.


\begin{thebibliography}{99}
\bibitem{ft2} Unfortunately, rather confusing terminology has been
used in the spin dynamics literature to describe this probability, including
persistence, survival, and non-zero-crossing probability. Further confusion
is caused by the fact that even in the surface dynamics problem, persistence
and survival become the same if one starts from the initial
condition of a ``flat surface'' (which cannot be achieved experimentally)
and defines ``survival'' with respect to the average surface which is also
``flat''.
\bibitem{ft3} We point out in this context that in
Ref.~\cite{us1} both of these probabilities were unfortunately
called persistence probability (with respect to the average position and
the initial position, respectively) causing perhaps some confusion.
\end{thebibliography}
\end{document}